%% file: ms.tex
\documentclass[sigconf,10pt]{acmart}
\def\BibTeX{{\rm B\kern-.05em{\sc i\kern-.025em b}\kern-.08emT\kern-.1667em\lower.7ex\hbox{E}\kern-.125emX}}
    
\acmSubmissionID{123}

\newcommand\blfootnote[1]{%
  \begingroup
  \renewcommand\thefootnote{}\footnote{#1}%
  \addtocounter{footnote}{-1}%
  \endgroup
}

\pdfoutput=1
\usepackage{booktabs}
\usepackage{xspace}
\usepackage{tabularx}
\usepackage{amssymb}
\usepackage{pifont}
\usepackage{subcaption}
\usepackage{enumitem}
\usepackage{ifthen}
\usepackage{soul}
\newboolean{publicversion}

\usepackage{titlesec}
\titlespacing*{\subsection}{5pt}{0.5\baselineskip}{2pt}

\setboolean{publicversion}{true}

\ifthenelse{\boolean{publicversion}}{
  \newcommand{\grumbler}[3]{}
}{
  \newcommand{\grumbler}[3]{\xspace\textcolor{#3}{\bf #1 #2}}
}

\newcommand{\cmark}{\ding{51}}%
\newcommand{\xmark}{\ding{55}}%

\ifthenelse{\boolean{publicversion}}{
  
}{
  
}
\ifthenelse{\boolean{publicversion}}{
  \newcommand{\cameditstwo}[1]{#1}
}{
  \newcommand{\cameditstwo}[1]{\textcolor{blue}{\bf #1}}
}
\ifthenelse{\boolean{publicversion}}{
  
}{
  
}
\ifthenelse{\boolean{publicversion}}{
  \newcommand{\rk}[1]{\xspace#1}
}{
  \newcommand{\rk}[1]{{\xspace\textcolor{blue}{\bf #1}}}
}

\newcommand{\ak}[1]{\grumbler{Aasheesh:}{#1}{purple}}

\newcommand{\todo}[1]{\grumbler{TODO}{#1}{red}}

\newcommand{\ra}[1]{\renewcommand{\arraystretch}{#1}}
\newcolumntype{R}[1]{>{\raggedleft\arraybackslash}p{#1}}

\input{macros}

\setcopyright{acmcopyright}
\begin{document}

%
\title{SplitFS: Reducing Software Overhead\\ in File
  Systems for Persistent Memory\\
}

\author{Rohan Kadekodi}
\affiliation{%
  \institution{University of Texas at Austin}
}

\author{Se Kwon Lee}
\affiliation{%
  \institution{University of Texas at Austin}
}

\author{Sanidhya Kashyap}
\affiliation{%
  \institution{Georgia Institute of Technology}
}

\author{Taesoo Kim}
\affiliation{%
  \institution{Georgia Institute of Technology}
}

\author{Aasheesh Kolli}
\affiliation{%
  \institution{Pennsylvania State University and VMware Research}
}

\author{Vijay Chidambaram}
\affiliation{%
  \institution{University of Texas at Austin and VMware Research}
  \vspace{3ex}
}

\renewcommand{\shortauthors}{R. Kadekodi et al.}

\input{abstract}

%
%


%

%
\copyrightyear{2019}
\acmYear{2019}
\acmConference[SOSP '19]{SOSP '19: Symposium on Operating Systems Principles}{October 27--30, 2019}{Huntsville, ON, Canada}
\acmBooktitle{SOSP '19: Symposium on Operating Systems Principles, October 27--30, 2019, Huntsville, ON, Canada}
\acmPrice{15.00}
\acmDOI{10.1145/3341301.3359631}
\acmISBN{978-1-4503-6873-5/19/10}

%
\settopmatter{printfolios=true, printacmref=true, printccs=true}
\maketitle
{
\blfootnote{Rohan Kadekodi and Se Kwon Lee are supported by SOSP 2019
student travel scholarships from the National Science Foundation.}
  \input{intro2}
  \input{background}

  \input{design2}

  \input{discussion}
  \input{eval}
  \input{related}

\input{conc}

  \input{ack}
}
             
\newpage

\bibliographystyle{ACM-Reference-Format}
\bibliography{all}

%

\end{document}

%% file: macros.tex
\newcommand{\mycaption}[2]{\caption{\textbf{#1}. {#2}}}
\newcommand{\sref}[1]{\S\ref{#1}}
\newcommand{\vheading}[1]{\vspace{0.05in}\noindent\textbf{#1}}

\newcommand{\fsync}{\texttt{fsync()}\xspace}
\newcommand{\etal}{\textit{et al.}\xspace}

\newcommand{\eg}{\textit{e.g.,}\xspace}

\newcommand{\myx}{$\times$\xspace}
\newcommand{\vtt}[1]{\texttt{#1}\xspace}

\newcommand{\syswrite}{\texttt{write()}\xspace}
\newcommand{\sysread}{\texttt{read()}\xspace}
\newcommand{\sysopen}{\texttt{open()}\xspace}
\newcommand{\sysclose}{\texttt{close()}\xspace}
\newcommand{\sysunlink}{\texttt{unlink()}\xspace}
\newcommand{\sysstat}{\texttt{stat()}\xspace}
\newcommand{\sysrename}{\texttt{rename()}\xspace}

\newcommand{\sysexecve}{\texttt{execve()}\xspace}
\newcommand{\sysmmap}{\texttt{mmap()}\xspace}
\newcommand{\sysfsync}{\texttt{fsync()}\xspace}

\newcommand{\mmap}{\texttt{mmap()}\xspace}

\newcommand{\mmaped}{\texttt{mmap}-ed\xspace}

\newcommand{\clflush}{\texttt{clflush}\xspace}

\newcommand{\sysname}{{\textsc{SplitFS}}\xspace}

\newcommand{\extdax}{ext4 DAX\xspace}

\newcommand{\sfence}{\texttt{sfence}\xspace}

\newcommand{\preload}{{\small\texttt{LD\_PRELOAD}}\xspace}

\definecolor{cadmiumgreen}{rgb}{0.0, 0.4, 0.24}
\definecolor{calpolypomonagreen}{rgb}{0.12, 0.3, 0.17}
\definecolor{cocoabrown}{rgb}{0.82, 0.41, 0.12}
\definecolor{cornellred}{rgb}{0.7, 0.11, 0.11}

\newcommand{\ictl}{\vtt{ioctl}\xspace}
\newcommand{\sysfork}{\vtt{fork()}\xspace}
\newcommand{\usplit}{{\textsc{U-Split}}\xspace}
\newcommand{\ksplit}{{\textsc{K-Split}}\xspace}
\newcommand{\mus}{$\mu$s\xspace}
\newcommand{\ns}{$ns$\xspace}

%% file: abstract.tex
\begin{abstract}
We present \sysname, a file system for persistent memory (PM) that
reduces software overhead significantly compared to state-of-the-art
PM file systems. \sysname presents a novel split of responsibilities
between a user-space library file system and an existing kernel PM
file system. The user-space library file system handles data
operations by intercepting POSIX calls, memory-mapping the underlying
file, and serving the read and overwrites using processor loads and
stores. Metadata operations are handled by the kernel PM file system
(ext4 DAX). \sysname introduces a new primitive termed relink to
efficiently support file appends and atomic data operations. \sysname
provides three consistency modes, which different applications can
choose from, without interfering with each other. \sysname reduces
software overhead by up-to 4\myx compared to the NOVA PM file system,
and 17\myx compared to ext4 DAX. On a number of micro-benchmarks and
applications such as the LevelDB key-value store running the YCSB
benchmark, \sysname increases application performance by
\cameditstwo{up to 2\myx compared to ext4 DAX and NOVA} while
providing similar consistency guarantees.
  
\end{abstract}

%% file: intro2.tex
\section{Introduction}
\label{sec-intro}

Persistent Memory (PM) is a new memory technology that was recently
introduced by Intel~\cite{3dxpoint}. PM will be placed on the memory
bus like DRAM and will be accessed via processor loads and stores. PM
has a unique performance profile: compared to DRAM, loads have
2--3.7\myx higher latency and $1/3^{rd}$ bandwidth, while stores have
the same latency but $1/6^{th}$ bandwidth~\cite{pm-arxiv}.
A single machine can be equipped with up to 6 TB of PM. Given its 
large capacity and low latency, an important use case for PM
will be acting as storage.

Traditional file systems add large overheads to each file-system 
operation, especially on
the write path. The overhead comes from performing expensive
operations on the critical path, including allocation, logging, and
updating multiple complex structures. The systems community has
proposed different architectures to reduce overhead. BPFS~\cite{bpfs},
PMFS~\cite{pmfs}, and NOVA~\cite{nova} redesign the in-kernel file
system from scratch to reduce overhead for file-system
operations. Aerie~\cite{aerie} advocates a user-space library file
system coupled with a slim kernel component that does coarse-grained
allocations. Strata~\cite{strata} proposes keeping the file system
entirely in user-space, dividing the system between a user-space
library file system and a user-space metadata server. Aerie and Strata
both seek to reduce overhead by not involving the kernel for most
file-system operations.

\input{tbl-mot}

Despite these efforts, file-system data operations, especially writes,
have significant overhead. For example, consider the common operation
of appending 4K blocks to a file (total 128 MB). It takes $671$ \ns to
write a 4 KB to PM; thus, if performing the append operation took a
total of $675$ \ns, the software overhead would be $4$
\ns. Table~\ref{tbl-mot} shows the software overhead on the append
operation on various PM file systems. We observe that there is still
significant overhead ($3.5-12.4$\myx) for file appends.

This paper presents \sysname, a PM file system that seeks to
  reduce software overhead via a novel \emph{split} architecture: a
  user-space library file system handles data operations while a
  kernel PM file system (ext4 DAX) handles metadata operations. 
  We refer to all file system operations that modify file metadata
    as \emph{metadata operations}. Such operations include \sysopen,
    \sysclose, and even file appends (since the file size is changed).
  The novelty of \sysname lies in how responsibilities are divided between the user-space and kernel components, and the semantics provided to applications. Unlike prior work like Aerie, which used the kernel only for coarse-grained operations, or Strata, where all operations are in user-space, \sysname routes \emph{all} metadata operations to the kernel. While FLEX~\cite{flex} invokes the kernel at a fine granularity like
  \sysname, it does not provide strong semantics such as synchronous,
  atomic operations to applications. At a high level, the \sysname
  architecture is based on the belief that if we can accelerate
  common-case data operations, it is worth paying a cost on the
  comparatively rarer metadata operations. This is in contrast with
  in-kernel file systems like NOVA which extensively modify the file
  system to optimize the metadata operations.

\sysname transparently reduces software overhead for reads and
overwrites by intercepting POSIX calls, memory mapping the underlying
file, and serving reads and overwrites via processor loads and
stores. \sysname optimizes file appends by introducing a new primitive
named \emph{relink} that minimizes both data copying and trapping into
the kernel. The application does not have to be rewritten in any way
to benefit from \sysname. \sysname reduces software overhead
by up-to 4\myx compared to NOVA and 17\myx compared to ext4 DAX.

Apart from lowering software overhead, the split architecture leads to
several benefits. First, instead of re-implementing file-system
functionality, \sysname can take advantage of the mature, well-tested
code in ext4 DAX for metadata operations. Second, the user-space
library file system in \sysname allows each application to run with
one of three consistency modes (\emph{POSIX, sync, strict}). We
observe that not all applications require the same guarantees; for
example, SQLite does not require the strong guarantees provided by
NOVA-strict, and gets 2.5\myx higher throughput on ext4 DAX and
\sysname-POSIX than on NOVA-strict owing to their weaker guarantees.
Applications running with different consistency modes do not interfere
with each other on \sysname.

\sysname introduces the relink primitive to optimize file appends and
atomic data operations. Relink logically and atomically moves a
contiguous extent from one file to another, without any physical data
movement. Relink is built on top of the \vtt{swap\_extents}
\vtt{ioctl} in ext4 DAX, and uses ext4 journaling to ensure the source
and destination files are modified atomically. Both file appends and
data overwrites in strict mode are redirected to a temporary PM file
we term the staging file. On \sysfsync, the data from the staging file
is relinked into the original file. Relink provides atomic data
operations without paging faults or data copying.

\sysname also introduces an optimized logging protocol. In strict
mode, all data and metadata operations in \sysname are atomic and
synchronous. \sysname achieves this by logging each operation. In the
common case, \sysname will write a single cache line
worth of data (64B), followed by one memory fence (\eg \sfence in
  x86 systems), for each operation; in contrast, NOVA writes at least
two cache lines and issues two fences. As a result of these
optimizations, \sysname logging is 4\myx faster than NOVA in the
critical path. Thanks to relink and optimized logging, atomic data
operations in \sysname are 2--6\myx faster than in
NOVA-strict, providing strong guarantees at low software overhead.

We evaluate \sysname using a number of micro-benchmarks, three
utilities (git, tar, rsync), two key-value stores (Redis, LevelDB),
and an embedded database (SQLite). Our evaluation on
\cameditstwo{Intel DC Persistent Memory} shows that \sysname, though
it is built on ext4 DAX, outperforms ext4 DAX by up-to 2\myx on many
workloads. \sysname outperforms NOVA by 10\%--2\myx (when providing
the same consistency guarantees) on LevelDB, Redis, and SQLite when
running benchmarks like YCSB and TPCC. \sysname also reduces total
amount of write IO by 2\myx compared to Strata on certain
workloads. On metadata-heavy workloads such as git and tar, \sysname
suffers a modest drop in performance (less than 15\%) compared to NOVA
and ext4 DAX.

\sysname is built on top of ext4 DAX; this is both a strength and a
weakness. Since \sysname routes all metadata operations through ext4
DAX, it suffers from the high software overhead and high write IO for
metadata operations. Despite these limitations, we believe \sysname
presents a useful new point in the spectrum of PM file-system
designs. ext4 DAX is a robust file system under active development;
its performance will improve with every Linux kernel version. \sysname
provides the best features of ext4 DAX while making up for its lack of
performance and strong consistency guarantees. 

This paper makes the following contributions:
\begin{itemize}
  \item A new architecture for PM file systems with a novel split of
    responsibilities between a user-space library file system and a
    kernel file system.
  \item The novel relink primitive that can be used to provide efficient
    appends and atomic data operations.
  \item The design and implementation of \sysname, based on the split
    architecture. We have made \sysname publicly available at \url{https://github.com/utsaslab/splitfs}.
  \item Experimental evidence demonstrating that \sysname outperforms
    state-of-the-art in-kernel and in-user-space PM file systems, on 
    a range of workloads. 

\end{itemize}  

%% file: tbl-mot.tex
\begin{table}[!tb]
  \small
  \centering
  \vspace{5pt}
  \begin{tabular}{@{}p{85pt}p{45pt}p{45pt}p{40pt}@{}}
    \toprule[1.2pt]
    File system &  Append Time (\ns) & Overhead (\ns) & Overhead (\%)\\
    \midrule
    ext4 DAX & 9002 & 8331 & 1241\% \\
    PMFS & 4150 & 3479 & 518\% \\
    NOVA-Strict & 3021 & 2350 & 350\% \\
    \midrule
    \sysname-Strict & 1251 & 580 & 86\% \\
    \sysname-POSIX & 1160 & 488 & 73\% \\
    \bottomrule[1.2pt]
  \end{tabular}
  \vspace{5pt} \mycaption{Software Overhead}{The table shows the
    software overhead of various PM file systems for appending a 4K
    block. It takes $671$ \ns to write 4KB to PM. Strict and POSIX
    indicate the guarantees offered by the file systems
    (\sref{sec-modes}).}
  \label{tbl-mot}
\end{table}

%% file: background.tex
\section{Background}
\label{sec-bkgd}

This section provides background on persistent memory (PM), PM file
systems, Direct Access, and memory mapping.


\subsection{Persistent Memory}

Persistent memory is a new memory technology that offers durability
and performance close to that of DRAM. PM can be attached on the
memory bus similar to DRAM, and would be accessed via processor loads
and stores. PM offers 8-byte atomic stores and they become persistent
as soon as they reach the PM controller~\cite{intel_adr}. There are 
two ways to ensure that stores become persistent: (i) using non-temporal
store instructions (e.g., \texttt{movnt} in x86) to bypass the cache
hierarchy and reach the PM controller or (ii) using a combination
of regular temporal store instructions and cache line flush instructions
(e.g., \texttt{clflush} or \texttt{clwb} in x86). 

Intel DC Persistent Memory is the first PM product that was made
commercially available in April 2019. Table~\ref{tbl-pm} lists the
performance characteristics of PM revealed in a report by Izraelevitz
\etal~\cite{pm-arxiv}. Compared to DRAM, PM has 3.7\myx higher latency
for random reads, 2\myx higher latency for sequential reads,
$1/3^{rd}$ read bandwidth, and close to $1/6^{th}$ write bandwidth.
Finally, PMs are expected to exhibit limited write endurance (about
$10^7$ write cycles~\cite{pm-wear}).

\input{tbl-pm}

\subsection{Direct Access (DAX) and Memory Mapping}
The Linux ext4 file system introduced a new mode called Direct Access
(DAX) to help users access PM~\cite{ext4-dax}. DAX file systems eschew the use of 
page caches and rely on memory mapping to provide low-latency access
to PM.

A memory map operation (performed via the \sysmmap system call) in
ext4 DAX maps one or more pages in the process virtual address space
to extents on PM. For example, consider virtual addresses \vtt{4K} to
\vtt{8K-1} are mapped to bytes \vtt{0} to \vtt{4K-1} on file \vtt{foo}
on ext4 DAX. Bytes \vtt{0} to \vtt{4K-1} in \vtt{foo} then correspond
to bytes \vtt{10*4K} to \vtt{11*4K -1} on PM. A store instruction to
virtual address \vtt{5000} would then translate to a store to byte
\vtt{40964} on PM. Thus, PM can be accessed via processor loads and
stores without the interference of software; the virtual memory
subsystem is in charge of translating virtual addresses into
corresponding physical addresses on PM.  

While DAX and \sysmmap provide low-latency access to PM, they do not
provide other features such as naming or atomicity for operations. The
application is forced to impose its own structure and semantics on the
raw bytes offered by \sysmmap. As a result, PM file systems still
provide useful features to applications and end users.

\subsection{PM File Systems}

Apart from ext4 DAX, researchers have developed a number of other PM
file systems such as SCMFS~\cite{scmfs}, BPFS~\cite{bpfs},
Aerie~\cite{aerie}, PMFS~\cite{pmfs}, NOVA~\cite{nova}, and
Strata~\cite{strata}. 
Only ext4 DAX, PMFS (now
deprecated), NOVA, and Strata are publicly available and supported by
modern Linux 4.x kernels.

These file systems make trade-offs between software overhead, amount
of write IO, and operation guarantees. NOVA provides strong guarantees
such as atomicity for file-system operations. PMFS provides slightly
weaker guarantees (data operations are not atomic), but as a result
obtains better performance on some workloads. Strata is a cross-media
file system which uses PM as one of its layers. Strata writes all data
to per-process private log, then coalesces the data and copies it to a
shared area for public access. For workloads dominated by operations
such as appends, Strata cannot coalesce the data effectively, and has
to write data twice: once to the private log, and once to the shared
area. This increases the PM wear-out by up to 2\myx. All these file
systems still suffer from significant overhead for write operations
(Table~\ref{tbl-mot}).

%% file: tbl-pm.tex
\begin{table}[!tb]
  \small
  \centering
  \begin{tabular}{@{}p{125pt}p{35pt}p{50pt}@{}}
    \toprule[1.2pt]
    Property & DRAM & Intel PM \\
    \midrule
    Sequential read latency (ns) &  81 & 169 (2.08\myx)\\
    Random read latency (ns) &  81 & 305 (3.76\myx)\\
    Store + flush + fence (ns) & 86 & 91 (1.05\myx)\\
    \midrule
    Read bandwidth (GB/s) & 120 & 39.4 (0.33\myx)\\
    Write bandwidth (GB/s) & 80 & 13.9 (0.17\myx)\\
    \bottomrule[1.2pt]
  \end{tabular}
  \vspace{5pt} \mycaption{PM Performance}{The table shows performance
    characteristics of DRAM, PM and the ratio of PM/DRAM, as reported by Izraelevitz et al.~\cite{pm-arxiv}.}
  \label{tbl-pm}
  \vspace{-15pt}
\end{table}

%% file: design2.tex
\section{SplitFS: Design and Implementation}
\label{sec-design}

We present the goals of \sysname, its three modes and their
guarantees. We present an overview of the design, describe how
different operations are handled, and discuss how \sysname provides
atomic operations at low overhead. We describe the implementation of
\sysname, and discuss its various tuning parameters. Finally, we
discuss how the design of \sysname affects security.

\subsection{Goals}
\vheading{Low software overhead}. \sysname aims to reduce software
  overhead for data operations, especially writes and appends.

  \vheading{Transparency}. \sysname does not require the application to
  be modified in any way to obtain lower software overhead and
  increased performance. 
  
  \vheading{Minimal data copying and write IO}. \sysname aims to reduce
  the number of writes made to PM.  \sysname aims to avoid copying
  data within the file system whenever possible. This both helps
  performance and reduces wear-out on PM. Minimizing writes is
  especially important when providing strong guarantees like atomic
  operations. 

  \vheading{Low implementation complexity}. \sysname aims to re-use
  existing software like ext4 DAX as much as possible, and reduce the
  amount of new code that must be written and maintained for \sysname.

  \vheading{Flexible guarantees}. \sysname aims to provide
  applications with a choice of crash-consistency guarantees to choose
  from. This is in contrast with PM file systems today, which provide
  all running applications with the same set of guarantees.
  

\subsection{\sysname Modes and Guarantees}
\label{sec-modes}

\sysname provides three different modes: POSIX, sync, and strict. Each
mode provides a different set of guarantees. Concurrent applications
can use different modes at the same time as they run on
\sysname. Across all modes, \sysname ensures the file system retains
its integrity across crashes.

\input{tbl-modes-2}

Table~\ref{tbl-modes} presents the three modes provided by
\sysname. Across all modes, appends are atomic in \sysname; if a
series of appends is followed by \sysfsync, the file will be
atomically appended on \sysfsync.

\vheading{POSIX mode}. In POSIX mode, \sysname provides metadata
consistency~\cite{Chidambaram+12-NoFS}, similar to ext4 DAX. The file
system will recover to a consistent state after a crash with respect
to its metadata. In this mode, overwrites are performed in-place and
are synchronous. Note that appends are not synchronous, and require an
\sysfsync to be persisted. However, \sysname in the POSIX mode
guarantees atomic appends, a property not provided by ext4 DAX.  This
mode slightly differs from the standard POSIX semantics: when a file
is accessed or modified, the file metadata will not immediately
reflect that.

\vheading{Sync mode}. \sysname ensures that on top of POSIX mode
guarantees, operations are also guaranteed to be synchronous. An
operation may be considered complete and persistent once the
corresponding call returns and applications do not need a subsequent
\fsync. Operations are not atomic in this mode; a crash may leave a
data operation partially completed. No additional crash recovery needs
to be performed by \sysname in this mode. This mode provides
  similar guarantees to PMFS as well as NOVA without data and metadata
  checksuming and with in-place updates; we term this NOVA
  configuration \emph{NOVA-Relaxed}.

\vheading{Strict mode}. \sysname ensures that on top of sync mode
guarantees, each operation is also atomic. This is a useful guarantee for applications;
editors can allow atomic changes to the file when the user saves the
file, and databases can remove logging and directly update the
database. This mode does not provide atomicity across system calls
though; so it cannot be used to update two files atomically together.
This mode provides similar guarantees to a NOVA
  configuration we term \emph{NOVA-Strict}: NOVA with copy-on-write
  updates, but without checksums enabled.

\input{tbl-tech}

\vheading{Visibility}. Apart from appends, all \sysname operations
become immediately visible to all other processes on the system. On
\sysfsync, appends are persisted and become visible to the rest of the
system. \sysname is unique in its visibility guarantees, and takes the
middle ground between ext4 DAX and NOVA where all operations are
immediately visible, and Strata where new files and data updates are
only visible to other processes after the digest operation. Immediate
visibility of changes to data and metadata combined with atomic,
synchronous guarantees removes the need for leases to coordinate
sharing; applications can share access to files as they would on any
other POSIX file system.

\input{fig-design}

\subsection{Overview}
We now provide an overview of the design of \sysname, and how it uses
various techniques to provide the outlined
guarantees. Table~\ref{tbl-tech} lists the different techniques and
the benefit each technique provides.

\vheading{Split architecture}. As shown in Figure~\ref{fig-design},
\sysname comprises of two major components, a user-space library
linked to the application called \usplit and a kernel file system
called \ksplit. \sysname services all data operations (\eg \sysread
and \syswrite calls) directly in user-space and routes metadata
operations (\eg \fsync, \sysopen, etc. ) to the kernel file system
underneath. File system crash-consistency is guaranteed at all
times. This approach is similar to Exokernel~\cite{exokernel} where
only the control operations are handled by the kernel and data
operations are handled in user-space.

\vheading{Collection of mmaps}. Reads and overwrites are handled by
\mmap-ing the surrounding 2 MB part of the file, and serving reads via
\vtt{memcpy} and writes via non-temporal stores (\vtt{movnt}
instructions).  A single logical file may have data present in
multiple physical files; for example, appends are first sent to a
staging file, and thus the file data is spread over the original file
and the staging file. \sysname uses a \emph{collection of memory-maps}
to handle this situation. Each file is associated with a number of
open \sysmmap calls over multiple physical files, and reads and
over-writes are routed appropriately.

\vheading{Staging}. \sysname uses temporary files called staging files
for both appends and atomic data operations. Appends are first routed
to a staging file, and are later relinked on \sysfsync. Similarly,
file overwrites in strict mode are also first sent to staging
files and later relinked to their appropriate files.

\vheading{Relink}. On an \fsync, all the staged appends of a file must
be moved to the target file; in strict mode, overwrites have to be
moved as well. One way to move the staged appends to the target file
is to allocate new blocks and then copy appended data to them.
However, this approach leads to write amplification and high overhead.
To avoid these unnecessary data copies, we developed a new primitive
called \emph{relink}. Relink logically moves PM blocks from the
staging file to the target file without incurring any copies.

Relink has the following signature: \texttt{\small relink(file1,
  offset1, file2, offset2, size)}. Relink atomically moves data from
\texttt{offset1} of \vtt{file1} to \vtt{offset2} of \vtt{file2}. If
\vtt{file2} already has data at \vtt{offset2}, existing data blocks
are de-allocated. Atomicity is ensured by wrapping the changes in a
ext4 journal transaction. Relink is a metadata operation, and does not
involve copying data when the involved \vtt{offsets} and \vtt{size}
are block aligned. When \vtt{offset1} or \vtt{offset2} happens to be
in the middle of a block, \sysname copies the partial data for that
block to \vtt{file2}, and performs a metadata-only relink for the rest
of the data. Given that \sysname is targeted at POSIX applications,
block writes and appends are often block-aligned by the
applications. Figure~\ref{fig-relink} illustrates the different steps
involved in the relink operation.

\input{fig-relink}

\vheading{Optimized logging}. In strict mode, \sysname guarantees
atomicity for all operations.  To provide atomicity, we employ an
\emph{Operation Log} and use logical redo logging to record the intent
of each operation. Each \usplit instance has its own operation
  log that is pre-allocated, \mmaped by \usplit, and written using
  non-temporal store instructions. We use the necessary memory fence
instructions to ensure that log entries persist in the correct
order. To reduce the overheads from logging, we ensure that in the
common case, per operation, we write one cache line (64B) worth of
data to PM and use a single memory fence (\sfence in x86) instruction
in the process. Operation log entries do not contain the file data
associated with the operation (\eg data being appended to a file),
instead they contain a logical pointer to the staging file where the
data is being held.

We employ a number of techniques to optimize logging. First, to
distinguish between valid and invalid or torn log entries, we
incorporate a 4B transactional checksum~\cite{ironfs} within the 64B
log entry. The use of checksum reduces the number of fence
instructions necessary to persist and validate a log entry from two to
one. Second, we maintain a tail for the log in DRAM and concurrent
threads use the tail as a synchronization variable. They use
compare-and-swap to atomically advance the tail and write to their
respective log entries concurrently. Third, during the initialization
of the operation log file, we zero it out. So, during crash recovery,
we identify all non-zero 64B aligned log entries as being potentially
valid and then use the checksum to identify any torn entries. The rest
are valid entries and are replayed. Replaying log entries is
idempotent, so replaying them multiple times on crashes is safe. We
employ a 128MB operation log file and if it becomes full, we
checkpoint the state of the application by calling \emph{relink()}
  on all the open files that have data in staging files. We then zero
out the log and reuse it. Finally, we designed our logging mechanism
such that all common case operations (\syswrite, \sysopen, etc.) can
be logged using a single 64B log entry while some uncommon operations,
like \sysrename, require multiple log entries.

Our logging protocol works well with the \sysname
  architecture. The tail of each \usplit log is maintained only in DRAM
  as it is not required for crash recovery. Valid log entries are
  instead identified using checksums. In contrast, file systems
  such as NOVA have a log per inode that resides on PM, whose tail is
  updated after each operation via expensive \clflush and \sfence
  operations.

\vheading{Providing Atomic Operations}. In strict mode, \sysname
provides synchronous, atomic operations. Atomicity is provided in an
efficient manner by the combination of staging files, relink, and
optimized logging.  Atomicity for data operations like overwrites is
achieved by redirecting them also to a staging file, similar to how
appends are performed.  \sysname logs these writes and appends to
record where the latest data resides in the event of a crash.  On
\sysfsync, \sysname relinks the data from the staging file to the
target file atomically. Once again, the data is written exactly once,
though \sysname provides the strong guarantee of atomic data
operations. Relink allows \sysname to implement a form of localized
copy-on-write. Due to the staging files being pre-allocated, locality
is preserved to an extent. \sysname logs metadata operations to ensure
they are atomic and synchronous. Optimized logging ensures that for
most operations exactly one cache line is written and one \vtt{sfence}
is issued for logging.

\subsection{Handling Reads, Overwrites, and Appends}

\vheading{Reads}. Reads consult the collection of mmaps to determine
where the most recent data for this offset is, since the data could
have been overwritten or appended (and thus in a staging file). If a
valid memory mapped region for the offsets being read exists in \usplit,
the read is serviced from the corresponding region.  If such a region
does not exist, then the 2 MB region surrounding the read offset is
first memory mapped, added to the the collection of mmaps, and then
the read operation is serviced using processor loads.

\vheading{Overwrites}. Similar to reads, if the target offset is
already memory mapped, then \usplit services the overwrite using
non-temporal store instructions. If the target offset is not memory
mapped, then the 2MB region surrounding the offset is first memory
mmaped, added to the collection of mmaps, and then the overwrite is
serviced. However, in strict mode, to guarantee atomicity, overwrites
are first redirected to a staging file (even if the offset is memory
mapped), then the operation is logged, and finally relinked on a
subsequent \sysfsync or \sysclose.

\vheading{Appends}. \sysname redirects all appends to a staging file,
and performs a relink on a subsequent \sysfsync or \sysclose. As
with overwrites, appends are performed with non-temporal writes and in
strict mode, \sysname also logs details of the append operation to
ensure atomicity.

\subsection{Implementation}

We implement \sysname as a combination of a user-space library file
system (9K lines of C code) and a small patch to ext4 DAX to add the
relink system call (500 lines of C code). \sysname supports 35 common
POSIX calls, such as \vtt{pwrite()}, \vtt{pread64()}, \vtt{fread()},
\vtt{readv()}, \vtt{ftruncate64()}, \vtt{openat()}, etc; we found that
supporting this set of calls is sufficient to support a
variety of applications and micro-benchmarks. Since PM file systems
PMFS and NOVA are supported by Linux kernel version 4.13, we modified
4.13 to support \sysname. We now present other details of our
implementation.

\vheading{Intercepting POSIX calls}. \sysname uses \preload to
intercept POSIX calls and either serve from user-space or route them
to the kernel after performing some book-keeping tasks.
Since \sysname intercepts calls at the POSIX
level in \vtt{glibc} rather than at the system call level, \sysname
has to intercept several variants of common system calls like
\vtt{write()}. 

\vheading{Relink}. We implement relink by leveraging an \vtt{ioctl}
provided by ext4 DAX. The \vtt{EXT4\_IOC\_MOVE\_EXT} \ictl swaps
extents between a source file and a destination file, and uses
journaling to perform this atomically. The \vtt{ioctl} also
de-allocates blocks in the target file if they are replaced by blocks
from the source file. By default, the \ictl also flushes the swapped
data in the target file; we modify the \ictl to only touch metadata,
without copying, moving, or persisting of data. We also ensure that
after the swap has happened, existing memory mappings of both source
and destination files are valid; this is vital to \sysname performance
as it avoids page faults. The \ictl requires blocks to be
  allocated at both source and destination files. To satisfy this
  requirement, when handling appends via relink, we allocate blocks at
  the destination file, swap extents from the staging file, and then
  de-allocate the blocks. This allows us to perform relink without
  using up extra space, and reduces implementation complexity at the
  cost of temporary allocation of data.

\vheading{Handling file open and close}. On file open, \sysname
performs \sysstat on the file and caches its attributes in user-space
to help handle later calls. When a file is closed, we do not clear its
cached information. When the file is unlinked, all cached metadata is
cleared, and if the file has been memory-mapped, it is un-mapped.  The
cached attributes are used to check file permissions on every
subsequent file operation (\eg \sysread) intercepted by \usplit.

\vheading{Handling fork}. Since \sysname uses a user-space library
file system, special care needs to be taken to handle \sysfork and
\sysexecve correctly. When \sysfork is called, \sysname is copied into the address
space of the new process (as part of copying the address space of the
parent process), so that the new process can continue to access
\sysname.

\vheading{Handling execve}. \sysexecve overwrites the address space,
but open file descriptors are expected to work after the call
completes. To handle this, \sysname does the following: before
executing \sysexecve, \sysname copies its in-memory data about open
files to a shared memory file on \vtt{/dev/shm}; the file name is the
process ID. After executing \sysexecve, \sysname checks the shared
memory device and copies information from the file if it exists.

\vheading{Handling dup}. When a file descriptor is duplicated, the
file offset is changed whenever operations are performed on either
file descriptor. \sysname handles by maintaining a single offset per
open file, and using pointers to this file in the file descriptor
maintained by \sysname. Thus, if two threads \vtt{dup} a file
descriptor and change the offset from either thread, \sysname ensures
both threads see the changes.

\vheading{Staging files}. \sysname pre-allocates staging files at
startup, creating 10 files each 160 MB in size. Whenever a staging
file is completely utilized, a background thread wakes up and creates
and pre-allocates a new staging file. This avoids the overhead of
creating staging files in the critical path. 

\vheading{Cache of memory-mappings}. \sysname caches all
memory-mappings its creates in its collection of memory mappings. A
memory-mapping is only discarded on \vtt{unlink()}. This reduces the
cost of setting up memory mappings in the critical path on read or
write.

\vheading{Multi-thread access}. \sysname uses a lock-free queue for
managing the staging files. It uses fine-grained reader-writer locks to
protect its in-memory metadata about open files, inodes, and
memory-mappings. 

\subsection{Tunable Parameters}

\sysname provides a number of tunable parameters that
can be set by application developers and users for each
\usplit instance. These parameters affect the performance of
\sysname.

\vheading{\mmap size}. \sysname supports a configurable
size of \mmap for handling overwrites and reads.  Currently,
     \sysname supports \mmap sizes ranging from 2MB to 512MB. The
     default size is 2 MB, allowing \sysname to employ huge pages
     while pre-populating the mappings.

     \vheading{Number of staging files at startup}. There are ten
     staging files at startup by default; when a staging file is used
     up, \sysname creates another staging file in the background. We
     experimentally found that having ten staging files provides a
     good balance between application performance and the
     initialization cost and space usage of staging files.

     \vheading{Size of the operation log}. The default size of the
     operation log is 128MB for each \usplit instance. Since all log
     entries consist of a single cacheline in the common case,
     \sysname can support up to 2M operations without clearing the log
     and re-initializing it. This helps applications with small bursts
     to achieve good performance while getting strong semantics.

\subsection{Security}
    
    \sysname does not expose any new security vulnerabilities as
    compared to an in-kernel file system.  All metadata operations are
    passed through to the kernel which performs security
    checks. \sysname does not allow a user to open, read, or write a
    file to which they previously did not have permissions. The
    \usplit instances are isolated from each other in separate
    processes; therefore applications cannot access the data of other
    applications while running on \sysname. Each \usplit instance only
    stores book-keeping information in DRAM for the files that the
    application already has access to. An application that uses
    \sysname may corrupt its own files, just as in an in-kernel file
    system.

%% file: tbl-modes-2.tex
\begin{table}[!tb]
  \small
  \centering
  \begin{tabular}{@{}p{22pt}p{20pt}p{25pt}p{35pt}p{35pt}p{55pt}@{}}
    \toprule[1.2pt]
    \emph{Mode} & \emph{Sync. Data Ops} & \emph{Atomic Data Ops} & \emph{Sync. Metadata Ops} & \emph{Atomic Metadata Ops} & \emph{Equivalent to} \\
    \midrule
    POSIX & \xmark & \xmark & \xmark & \cmark & ext4-DAX \\
    \midrule
    sync & \cmark & \xmark & \cmark & \cmark & Nova-Relaxed, PMFS \\
    \midrule
    strict & \cmark & \cmark & \cmark & \cmark & NOVA-Strict, Strata \\
    \bottomrule[1.2pt]
  \end{tabular}
  \vspace{5pt} \mycaption{\sysname modes}{The table shows the three
    modes of \sysname, the guarantees provided by each
    mode, and list current file systems which provide the same
    guarantees.}
  \label{tbl-modes}
\end{table}

%% file: tbl-tech.tex
\begin{table}[!t]
  \small
  \centering
  \begin{tabular}{@{}p{110pt}p{120pt}@{}}
    \toprule[1.2pt]
    \emph{Technique} & \emph{Benefit}\\
    \midrule
    Split architecture & Low-overhead data operations, correct
    metadata operations \\
    \midrule
    Collection of memory-mmaps & Low-overhead data operations in the
    presence of updates and appends \\
    \midrule
    Relink + Staging & Optimized appends, atomic data operations, low write
    amplification \\
    \midrule
    Optimized operation logging & Atomic operations, low write
    amplification \\
    \bottomrule[1.2pt]
  \end{tabular}
  \vspace{5pt} \mycaption{Techniques}{The table lists each main
    technique used in \sysname along with the benefit it provides. The
  techniques work together to enable \sysname to provide strong
  guarantees at low software overhead.}
  \label{tbl-tech}
\end{table}

%% file: fig-design.tex
\begin{figure}
\centering
\includegraphics[width=0.49\textwidth]{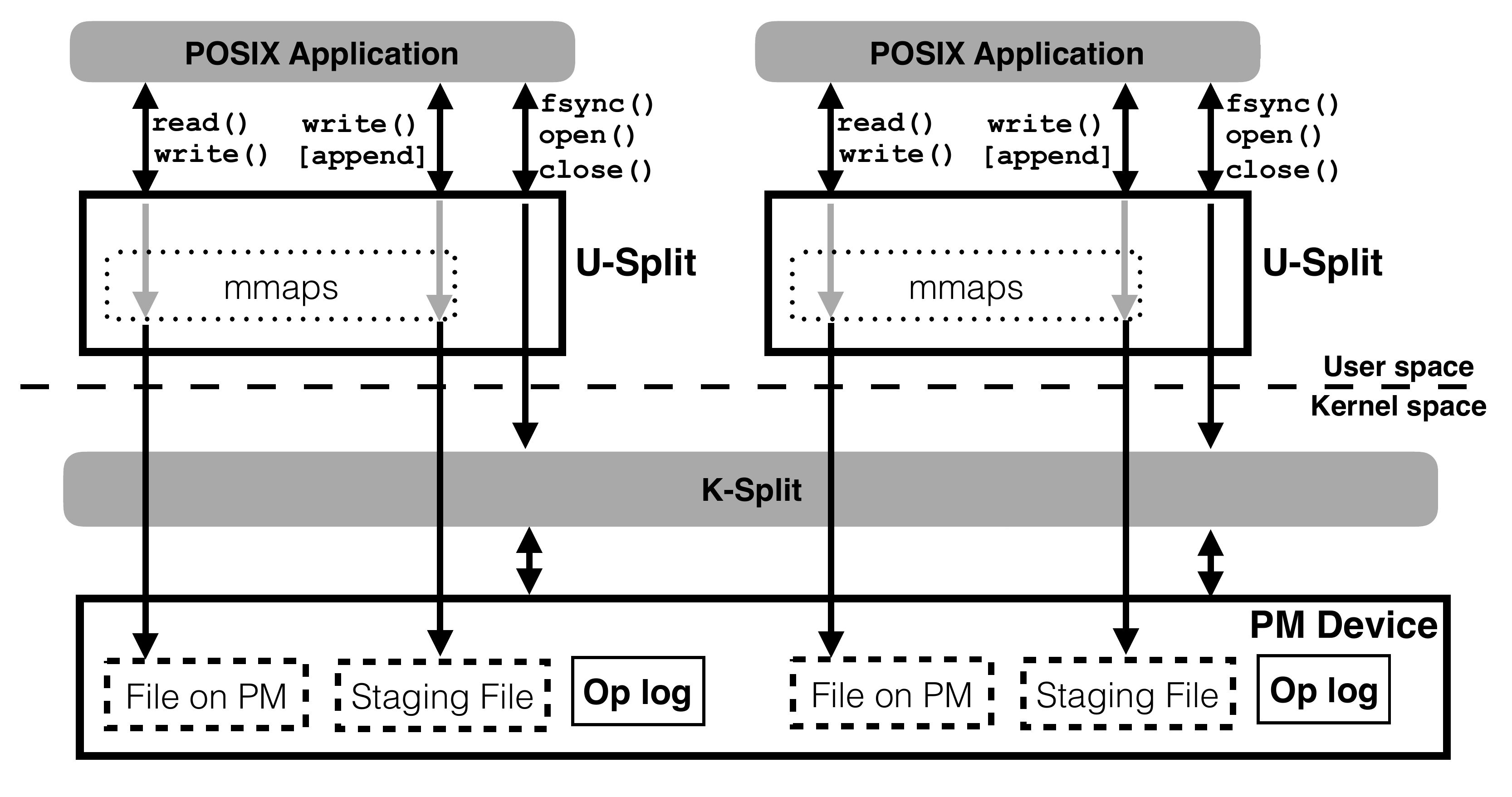}
\mycaption{\sysname Overview}{The figure provides an overview of how
  \sysname works. Read and write operations are transformed into loads
  and stores on the memory-mapped file. Append operations are staged in 
  a staging file and \emph{relinked} on \fsync. Other metadata POSIX 
  calls like \texttt{open()}, \texttt{close()}, etc. are passed through
  to the in-kernel PM file system. Note that loads and stores do not 
  incur the overhead of trapping into the kernel. 
}
\label{fig-design}
\end{figure}

%% file: fig-relink.tex
\begin{figure}
\centering
\includegraphics[width=0.47\textwidth]{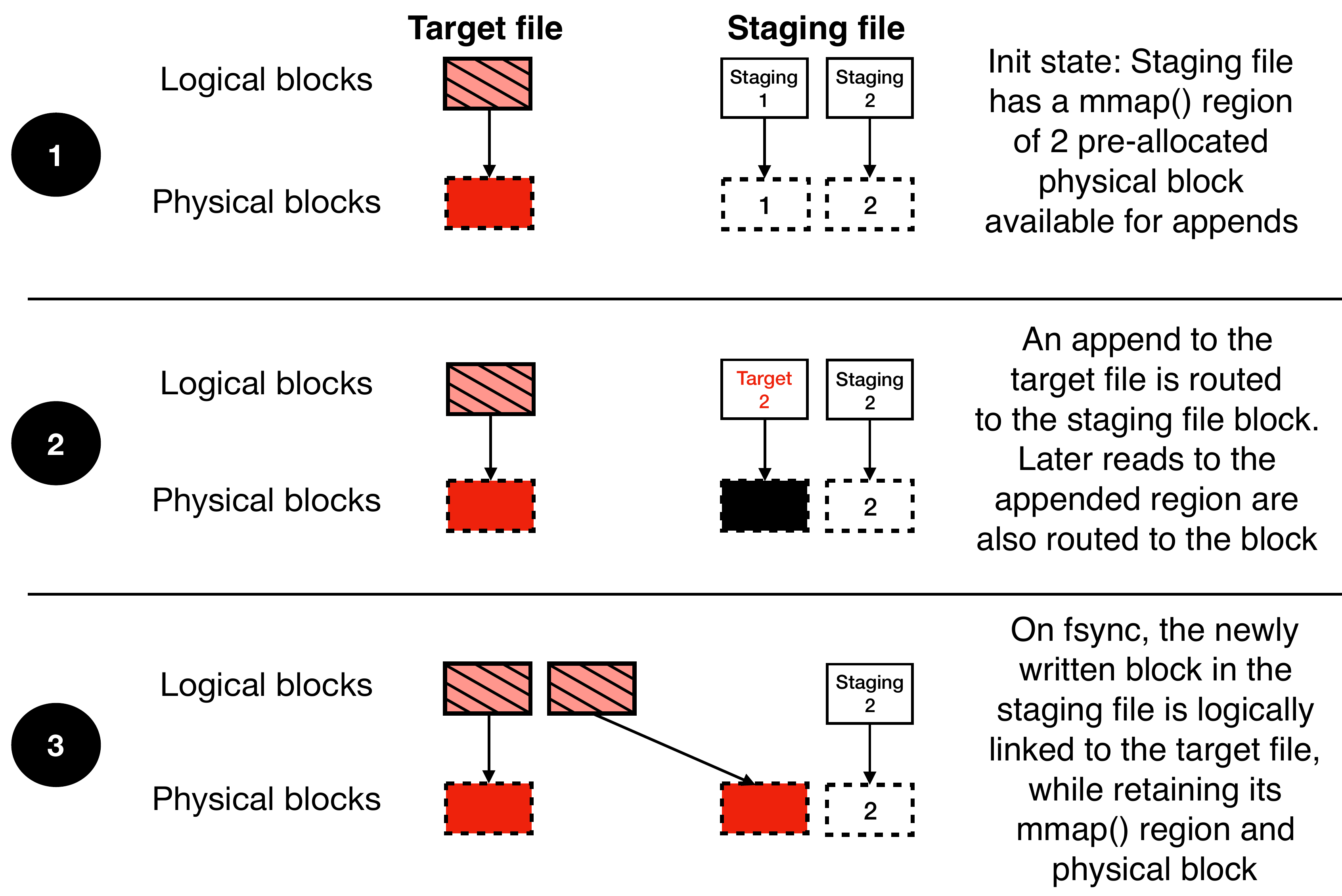}
\mycaption{\emph{relink} steps}{This figure provides an overview of
  the steps involved while performing a relink operation. First,
  appends to a target file are routed to pre-allocated blocks in the
  staging file and subsequently on an \sysfsync, they are \emph{relinked} into
  the target file while retaining existing memory-mapped regions.}
\label{fig-relink}
\end{figure}

%% file: discussion.tex
\section{Discussion}
\label{sec-discussion}

We reflect on our experiences building \sysname, describe problems we
encountered, how we solved them, and surprising insights that we
discovered.

\vheading{Page faults lead to significant cost}. \sysname memory maps
files before accessing them, and uses \vtt{MAP\_POPULATE} to pre-fault
all pages so that later reads and writes do not incur page-fault
latency. As a result, we find that a significant portion of the time
for \sysopen is consumed by page faults. While the latency of device
IO usually dominates page fault cost in storage systems based on solid
state drives or magnetic hard drives, the low latency of persistent
memory highlights the cost of page faults.

\vheading{Huge pages are fragile}. A natural way of minimizing page
faults is to use 2 MB huge pages. However, we found huge pages fragile
and hard to use. Setting up a huge-page mapping in the Linux kernel
requires a number of conditions. First, the virtual address must be 2
MB aligned. Second, the physical address on PM must be 2 MB
aligned. As a result, fragmentation in either the virtual address
space or the physical PM prevents huge pages from being created. For
most workloads, after a few thousand files were created and deleted,
fragmenting PM, we found it impossible to create any new huge
pages. Our collection-of-mappings technique sidesteps this problem by
creating huge pages at the beginning of the workload, and reusing them
to serve reads and writes. Without huge pages, we observed read
performance dropping by 50\% in many workloads. We believe this is a
fundamental problem that must be tackled since huge pages are crucial
for accessing large quantities of PM.

\vheading{Avoiding work in the critical path is important}. Finally,
we found that a general design technique that proved crucial for
\sysname is simplifying the critical path. We pre-allocate wherever
possible, and use a background thread to perform pre-allocation in the
background. Similarly, we pre-fault memory mappings, and use a cache
to re-use memory mappings as much as possible. \sysname rarely
performs heavyweight work in the critical path of a data
operation. Similarly, even in strict mode, \sysname optimizes logging,
trading off shorter recovery time for a simple, low overhead logging
protocol. We believe this design principle will be useful for other
systems designed for PM.

\vheading{Staging writes in DRAM}. An alternate design that we tried
was staging writes in DRAM instead of on PM. While DRAM staging files
incur less allocation costs than PM staging files, we
found that the cost of copying data from DRAM to PM on \sysfsync
overshadowed the benefit of staging data in DRAM. In general, DRAM
buffering is less useful in PM systems because PM and DRAM performances are similar.

\vheading{Legacy applications need to be rewritten to take maximum
  advantage of PM}. We observe that the applications we evaluate such
as LevelDB spent a significant portion of their time ($60-80$\%)
performing POSIX calls on current PM file systems. \sysname is able to
reduce this percentage down to 46-50\%, but further reduction in
software overhead will have negligible impact on application runtime
since the majority of the time is spent on application
code. Applications would need to be rewritten from scratch to use
libraries like \vtt{libpmem} that exclusively operate on data
structures in \sysmmap to take further advantage of PM.

%% file: eval.tex
\section{Evaluation}
\label{sec-eval}

In this section, we use a number of microbenchmarks and applications
to evaluate \sysname in relation to state-of-the-art PM filesystems
like \extdax, NOVA, and PMFS. While comparing these different file
systems, we seek to answer the following questions:

\begin{itemize}
[topsep=2pt,itemsep=2pt,parsep=2pt,before=\vspace{10pt},after=\vspace{10pt}]
	\item How does \sysname affect the performance of different 
	system calls as compared to \extdax?
	(\S\ref{sec-sys-call})
	\item How do the different techniques employed in \sysname
	contribute to overall performance? (\S\ref{sec-factor})
	\item How does \sysname compare to other file systems for
	different PM access patterns? (\S\ref{sec-micro})
	\item Does \sysname reduce file-system software overhead as
	compared to other PM file systems? (\S\ref{sec-sw-ovhd})
	\item How does \sysname compare to other file systems for
      real-world applications? (\S\ref{sec-real-di-apps} \&
      \S\ref{sec-real-mh-apps})
	\item What are the compute and storage overheads incurred when using \sysname? (\S\ref{sec-resource})
\end{itemize}

We first briefly describe our experimental methodology (\S\ref{sec-expsetup} \& \S\ref{sec-workloads}) before addressing each of the above questions.

\subsection{Experimental Setup}
\label{sec-expsetup}

\cameditstwo{We evaluate the performance of \sysname against other PM file systems on Intel Optane DC Persistent Memory Module (PMM). The experiments are performed on a 2-socket, 96-core machine with 768 GB PMM, 375 GB DRAM, and 32 MB Last Level Cache (LLC).} We run all evaluated
file systems on the 4.13 version of the Linux kernel (Ubuntu
16.04). We run each experiment multiple times and report the mean. In
all cases, the standard deviation was less than five percent of the
mean, and the experiments could be reliably repeated.

\subsection{Workloads}
\label{sec-workloads}

\input{tbl-app-desc.tex}

We used two key-value stores (Redis, LevelDB), an embedded database
(SQLite), and three utilities (tar, git, rsync) to evaluate the
performance of \sysname. Table~\ref{tbl-apps-desc} lists the
applications and their characteristics.

\vheading{TPC-C on SQLite}. TPC-C is an online transaction processing
benchmark. It has five different types of transactions each with
different ratios of reads and writes. We run SQLite v3.23.1 with
\sysname, and measured the performance of TPC-C on SQLite in the
Write-Ahead-Logging (WAL) mode.

\vheading{YCSB on LevelDB}. The Yahoo Cloud Serving
Benchmark~\cite{cooper2010benchmarking} has six different key-value
store benchmarks, each with different read/write ratios. We run the
YCSB workloads on the LevelDB key-value stores. We set the
\vtt{sstable} size to 64 MB as recommended in Facebook's tuning
guide~\cite{rocksdb-tuning}.

\vheading{Redis}. We set 1M key-value pairs in Redis~\cite{redis}, an
in-memory key-value store. We ran Redis in the Append-Only-File mode,
where it logs updates to the database in a file and performs \fsync on
the file every second.

\vheading{Utilities}. We also evaluated the performance of \sysname
for tar, git, and rsync. With git, we measured the time taken for
\vtt{git add} and \vtt{git commit} of all files in the Linux kernel
ten times.  With rsync, we copy a 7 GB dataset of 1200 files with
characteristics similar to backup datasets~\cite{backup} from one PM
location to another. With tar, we compressed the Linux kernel 4.18
along with the files from the backup dataset.

\subsection{Correctness and recovery}

\vheading{Correctness}. First, to validate the functional correctness
of \sysname we run various micro-benchmarks and real-world
applications and compare the resulting file-system state to the ones
obtained with \extdax. We observe that the file-system states obtained
with \extdax and \sysname are equivalent, validating how \sysname
  handles POSIX calls in its user-space library file system.

\vheading{Recovery times}. Crash recovery in POSIX and sync modes of
\sysname do not require anything beyond allowing the underlying ext4
DAX file system to recover. In strict mode however, all valid log
entries in the operation log need to be replayed on top of \extdax
recovery.  This additional log replay time depends on the number and
type of valid log entries in the log. To estimate the additional time
needed for recovery, we crash our real-world workloads at random
points in their execution and measure the log replay time. In our
crash experiments, the maximum number of log entries to be replayed
was 18,000 and that took about 3 seconds on emulated PM (emulation
details in \sref{sec-strata}). In a worst-case micro-benchmark where
we perform cache-line sized writes and crash with 2M (128MB of data)
valid log entries, we observed a log replay time of 6 seconds on
emulated PM.

\subsection{\sysname system call overheads}
\label{sec-sys-call}

The central premise of \sysname is that it is a good trade-off to
accelerate data operations at the expense of metadata operations.
Since data operations are more prevelant, this optimization improves
overall application performance.  To validate this premise, we
construct a micro-benchmark similar to FileBench
Varmail~\cite{tarasov2016filebench} that issues a variety of data and
metadata operations. The micro-benchmark first creates and appends
16KB to a file (as four appends, each followed by an \fsync), closes
it, opens it again, read the whole file as one read call, closes it,
then opens and closes the file once more, and finally deletes the
file. The multiple open and close calls were introduced to account for
the fact that their latency varies over time. Opening a file for the
first time takes longer than opening a file that we recently closed,
due to file metadata caching inside
\usplit. Table~\ref{tbl-micro-syscall} shows the latencies we observed
for different system calls and they are reported for all the three
modes provided by \sysname and for \extdax on which \sysname was
built.

\input{tbl-micro-syscall}

We make three observations based on these results. First, data
operations on \sysname are significantly faster than on
\extdax. Writes especially are 3--4\myx faster. Second, metadata
operations (\eg \sysopen, \sysclose, etc.) are slower on \sysname than
on \extdax, as \sysname has to setup its own data structures in
addition to performing the operation on \extdax. \rk {In \sysname, \sysunlink is an expensive operation because the file mappings that are created for serving reads and overwrites need to be unmapped in the \sysunlink wrapper.} Third, as the
consistency guarantees provided by \sysname get stronger, the syscall
latency generally increases. This increase can be attributed to more
work \sysname has to do (\eg logging in strict mode) for each system
call to provide stronger guarantees. Overall, \sysname achieves its
objective of accelerating data operations albeit at the expense of
metadata operations.

\subsection{\sysname performance breakdown}
\label{sec-factor}

\input{fig-factor}

We examine how the various techniques employed by \sysname contribute
to overall performance.  We use two write-intensive microbenchmarks:
sequential 4KB overwrites and 4KB appends. An \fsync is issued every
ten operations.  Figure~\ref{fig-factor} shows how individual
techniques introduced one after the other improve performance.

\vheading{Sequential overwrites}. \sysname increases sequential
overwrite performance by more than 2\myx compared to ext4 DAX since overwrites are served from
user-space via processor stores. However, further optimizations like
handling appends using staging files and relink have negligible impact
on this workload as it does not issue any file append operations.

\vheading{Appends}. The split architecture does not accelerate appends since without staging files or relink all appends go to \extdax as they are metadata operations.  Just introducing staging files to buffer appends improves performance by about 2\myx. In this setting, even though appends are serviced in user-space, overall performance is bogged down by expensive data copy operations on \fsync.  Introducing the relink primitive to this setting eliminates data copies and increases application throughput by 5\myx.

\subsection{Performance on different IO patterns}
\label{sec-micro}

\input{fig-micro-all}

To understand the relative merits of different PM file systems, we
compare their performance on microbenchmarks performing different file
IO patterns: sequential reads, random reads, sequential writes, random
writes, and appends. Each benchmark reads/writes an entire 128MB file
in 4KB operations. We compare file systems providing the same
guarantees: \sysname-POSIX with \extdax, \sysname-sync with PMFS, and
\sysname-strict with Nova-strict and
Strata. Figure~\ref{fig-micro-all} captures the performance of these
file systems for the different micro-benchmarks.

\vheading{POSIX mode}. \sysname is able to reduce the execution times
of \extdax by at least 27\% and as much as 7.85\myx (sequential reads and
appends respectively).  Read-heavy workloads present fewer improvement
opportunities for \sysname as file read paths in the kernel are
optimized in modern PM file systems. However, write paths are much
more complex and longer, especially for appends. So, servicing a write
in user-space has a higher payoff than servicing a read, an
observation we already made in Table~\ref{tbl-micro-syscall}.

\vheading{Sync mode}. Compared to PMFS, \sysname improves the
performance for write workloads (by as much as 2.89\myx) and increases performance for read workloads (by as much
as 56\%). Similar to \extdax, \sysname's ability to not incur
expensive write system calls translates to its superior performance
for the write workloads.

\vheading{Strict mode}. NOVA, Strata, and \sysname in this mode
provide atomicity guarantees to all operations and perform the
necessary logging. As can be expected, the overheads of logging result
in reduced performance compared to file systems in other
modes. Overall, \sysname improves the performance over NOVA by up to 5.8\myx
on the random writes workload. This improvement stems from \sysname's
superior logging which incurs half the number of log writes and fence
operations than NOVA. 

\subsection{Reducing software overhead}
\label{sec-sw-ovhd}

\input{fig-sw-ovhd-apps}

The central premise of \sysname is that it is possible to accelerate
applications by reducing file system software overhead. We define
file-system software overhead as the time taken to service a
file-system call minus the time spent actually accessing data on the
PM device.  For example, if a system call takes 100 \mus to be serviced,
of which only 25 \mus were spent read or writing to PM, then we say that
the software overhead is 75 \mus. To provide another example, for
appending 4 KB (which takes 10 \mus to write to PM), if file system A
writes 10 metadata items (incurring 100 \mus) while file system B
writes two metadata items (incurring 20 \mus), file-system B will have
lower overhead. In addition to avoiding kernel traps of system calls,
the different techniques discussed in \S\ref{sec-design} help \sysname
reduce its software overhead. Minimizing software overhead allows
applications to fully leverage PMs.

Figure~\ref{fig-sw-ovhd-apps} \cameditstwo{highlights the relative software
overheads incurred by different file systems compared to \sysname
providing the same level of guarantees. We present results for three
write-heavy workloads, LevelDB running YCSB Load A and Run A, and
SQLite running TPCC. \extdax and NOVA (in relaxed mode) suffer the
highest relative software overheads, up to 3.6\myx and 7.4\myx
respectively. NOVA-Relaxed incurs the highest software overhead for TPCC because it has to update the per-inode logical log entries on overwrites before updating the data in-place. On the other hand, \sysname-sync can directly perform in-place data updates, and thus has significantly lower software overhead. PMFS suffers the lowest relative software overhead,
capping off at 1.9\myx for YCSB Load A and Run A. Overall, \sysname incurs the lowest
software overhead.}

\subsection{Performance on data-intensive workloads}
\label{sec-real-di-apps}
\label{sec-strata}

\input{fig-real-apps}

Figure~\ref{fig-real-apps} summarizes the performance of various
applications on different file systems. The
performance metric we use for these data intensive workloads (LevelDB
with YCSB, Redis with 100\% writes, and SQLite with TPCC) is
throughput measured in KOps/s. For each mode of consistency guarantee
(POSIX, sync, and strict), we compare \sysname to state-of-the-art PM file systems. We report the absolute
performance for the baseline file system in each category and
relative throughput for \sysname. Despite our
  best efforts, we were not able to run Strata on these large applications; other researchers have also reported problems
  in evaluating Strata~\cite{flex}. We evaluated Strata
  with a smaller-scale YCSB workload using a 20GB private log.

Overall, \sysname outperforms other PM file systems (when providing
similar consistency guarantees) on all data-intensive workloads by as
much as \cameditstwo{2.70\myx}. We next present a breakdown of these numbers for
different guarantees.

\vheading{POSIX mode}. \cameditstwo{\sysname outperforms \extdax in all
workloads. Write-heavy workloads like RunA (2\myx), LoadA (89\%), LoadE
(91\%), Redis (27\%), etc.  benefit the most with \sysname. \sysname
speeds up writes and appends the most, so write-heavy workloads
benefit the most from \sysname. \sysname outperforms \extdax on
read-dominated workloads, but the margin of improvement is lower.}

\vheading{Sync and strict mode}. \cameditstwo{\sysname outperforms sync-mode file
systems PMFS and NOVA (relaxed) and strict-mode file system NOVA
(strict) for all the data intensive workloads. Once again, its the
write-heavy workloads that show the biggest boost in performance. For
example, \sysname in sync mode outperforms NOVA (relaxed) and PMFS by
2\myx and 30\% on RunA and in strict mode outperforms NOVA (strict) by
2\myx.  Ready-heavy workloads on the other hand do not show much
improvement in performance.}

\input{tbl-splitfs-strata.tex}

\vheading{Comparison with Strata}. We were able to reliably evaluate
Strata (employing a 20 GB private log) using LevelDB running
smaller-scale YCSB workloads (1M records, and 1M ops for workloads
A--D and F, 500K ops for workload E). We were unable to run Strata on
Intel DC Persistent Memory. Hence, we use DRAM to emulate PM. We
employ the same PM emulation framework used by Strata. We inject a
delay of 220ns on every \sysread system call, to emulate the access
latencies of the PM hardware. We do not add this fixed 220ns delay for
writes, because writes do not go straight to PM in the critical path,
but only to the memory controller. We add bandwidth-modeling delays
for reads as well as writes to emulate a memory device with $1/3^{rd}$
the bandwidth of DRAM, an expected characteristic of
PMs~\cite{pm-arxiv}. While this emulation approach is far from
perfect, we observe that the resulting memory access characteristics
are inline with the expected behavior of PMs~\cite{strata}. \sysname
outperforms Strata on all workloads, by 1.72\myx--2.25\myx as shown in
Table~\ref{tbl-splitfs-strata}.

\subsection{Performance on metadata-heavy workloads}
\label{sec-real-mh-apps}

Fig~\ref{fig-real-apps} compares the performance of \sysname with
other PM file systems (we only show the best performing PM file
system) on metadata-heavy workloads like git, tar, and rsync. These
metadata-heavy workloads do not present many opportunities for
\sysname to service system calls in userspace and in turn slow
metadata operations down due to the additional bookkeeping performed
by \sysname. These workloads represent the worst case scenarios for
\sysname. The maximum overhead experienced by \sysname is 13\%.

\subsection{Resource Consumption}
\label{sec-resource}

\sysname consumes memory for its file-related metadata (\eg, to keep
track of open file descriptors, staging files used). It also
additionally consumes CPU time to execute background threads that help
with metadata management and to move some expensive tasks off the
application's critical path.  

\vheading{Memory usage}. \sysname using a maximum of 100MB to maintain
its own metadata to help track different files, the mappings between
file offsets and \mmap-ed regions, etc. In strict mode, \sysname
additionally uses 40MB to maintain data structures to provide
atomicity guarantees.

\vheading{CPU utilization}.  \sysname uses a background thread to
handle various deferred tasks (\eg, stage file allocation, file
closures). This thread utilizes one physical thread
of the machine, occasionally increasing CPU consumption by 100\%.

%% file: tbl-app-desc.tex
\begin{table}[!t]
  \small
  \centering
  \ra{1.3}
  \begin{tabular}{@{}ll@{}}
    \toprule[1.2pt]
    Application & Description\\
    \midrule
    TPC-C~\cite{council2001tpc} on SQLite~\cite{pppsqlite} & Online transaction processing \\
    YCSB~\cite{cooper2010benchmarking} on LevelDB~\cite{pppleveldb} & Data retreival \& maintenance  \\
    Set in Redis~\cite{redis} & In-memory data structure store \\
    Git & Popular version control software \\
    Tar & Linux utility for data compression \\ 
    Rsync & Linux utility for data copy \\
    \bottomrule[1.2pt]
  \end{tabular}
  \vspace{5pt} \mycaption{Applications used in evaluation}{The table provides
    a brief description of the real-world applications we use to
    evaluate PM file systems.}
  \label{tbl-apps-desc}
  \vspace{-30pt}
\end{table}

%% file: tbl-micro-syscall.tex
\begin{table}[!t]
  \centering
  \ra{1.3}
  \begin{tabular}{@{}lrrrr@{}}
    \toprule[1.2pt]
    System call & Strict &  Sync & POSIX & \extdax \\
    \midrule
    open & 2.09 & 2.08 & 1.82 & 1.54 \\
    close & 0.78 & 0.69 & 0.69 & 0.34 \\
    append & 3.14 & 3.09 & 2.84 & 11.05\\
    fsync & 6.85 & 6.80 & 6.80 & 28.98 \\
    read & 4.57 & 4.53 & 4.53 & 5.04 \\
    unlink & 14.60 & 13.56 & 14.33 & 8.60 \\
    \bottomrule[1.2pt]
  \end{tabular}
  \vspace{5pt} \mycaption{\sysname system call overheads}{The table compares the
    latency (in us) of different system calls for various modes of 
    \sysname and \extdax.}
  \label{tbl-micro-syscall}
\end{table}

%% file: fig-factor.tex
\begin{figure}
\centering
\includegraphics[width=0.3\textwidth]{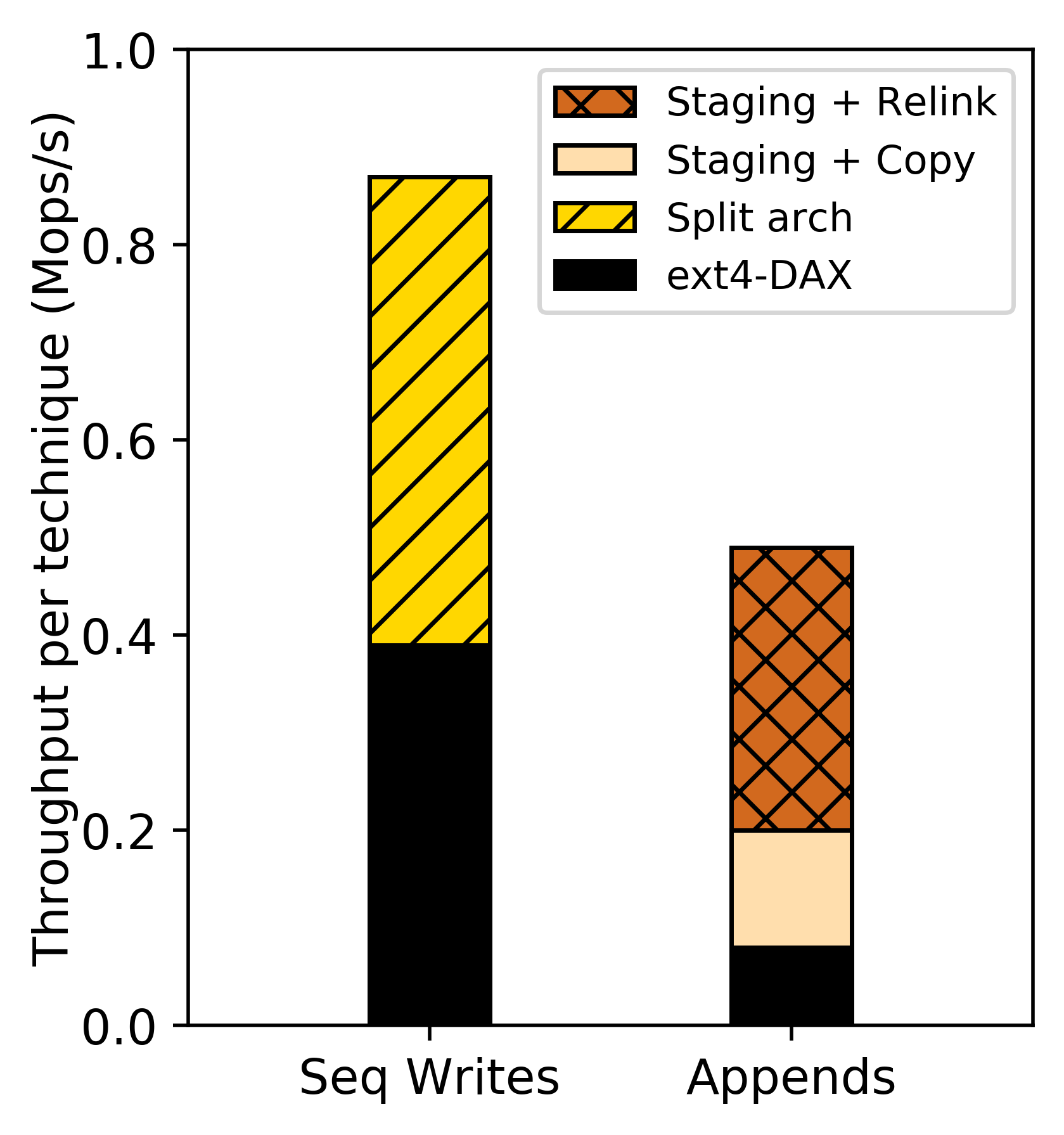}
\vspace{-10pt}
\mycaption{\sysname techniques contributions}{This figure shows
the contributions of different techniques to overall performance.
We compare the relative merits of these techniques using two
write intensive microbenchmarks; sequential overwrites and appends.}
\label{fig-factor}
\end{figure}

%% file: fig-micro-all.tex
\begin{figure}
\centering
\includegraphics[width=0.5\textwidth]{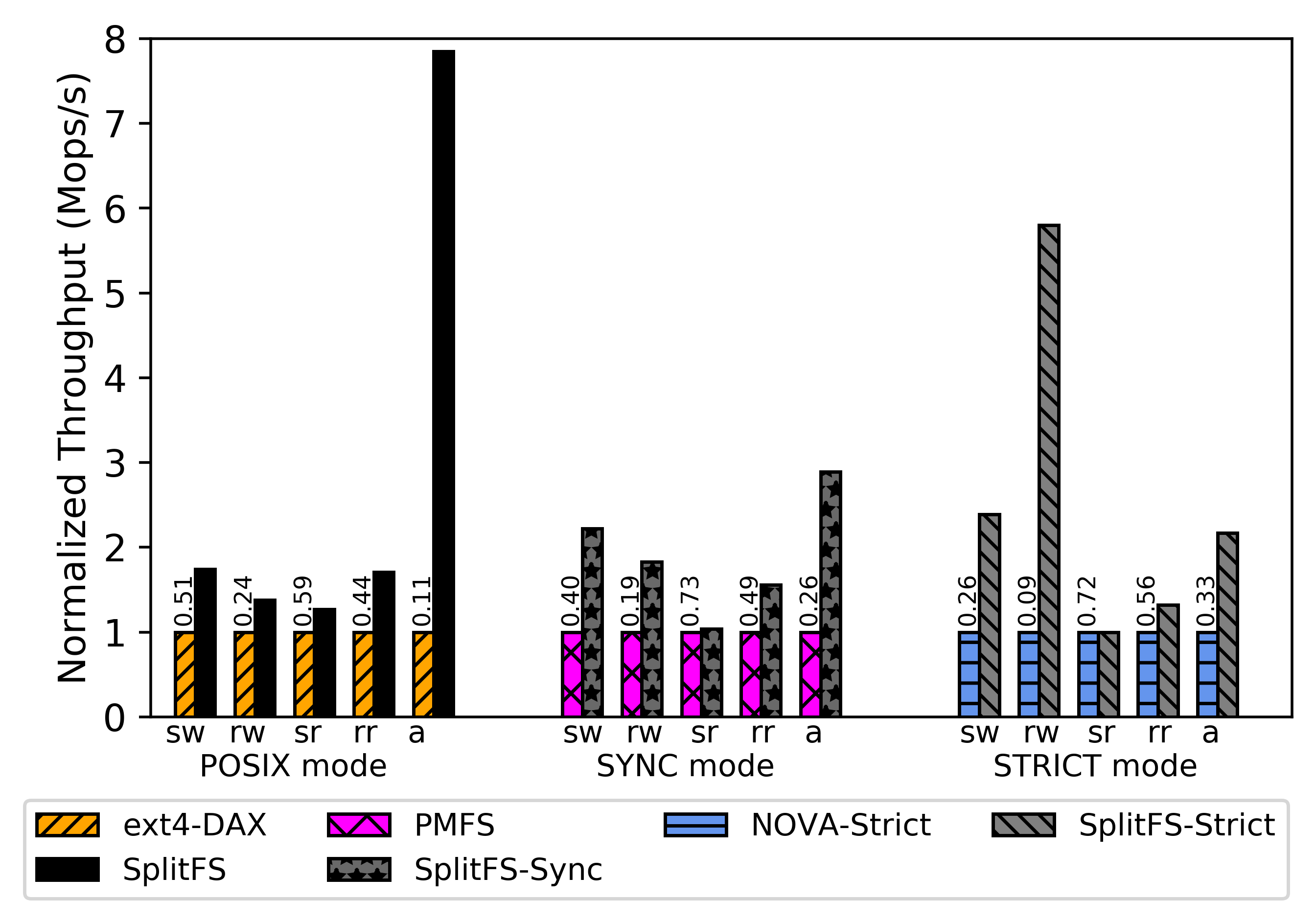}
\vspace{-20pt}
\mycaption{Performance on different IO patterns}{This figure compares
\sysname with the state-of-the-art PM file systems
in their respective modes using micrbenchmarks that 
perform five different kinds of file access patters. The y-axis is
throughput normalized to ext4 DAX in POSIX mode, PMFS in sync mode,
and NOVA-Strict in Strict mode (higher is better). The absolute
throughput numbers in Mops/s are given over the baseline in each
group.}
\label{fig-micro-all}
\end{figure}

%% file: fig-sw-ovhd-apps.tex
\begin{figure}
\centering
\includegraphics[width=0.5\textwidth]{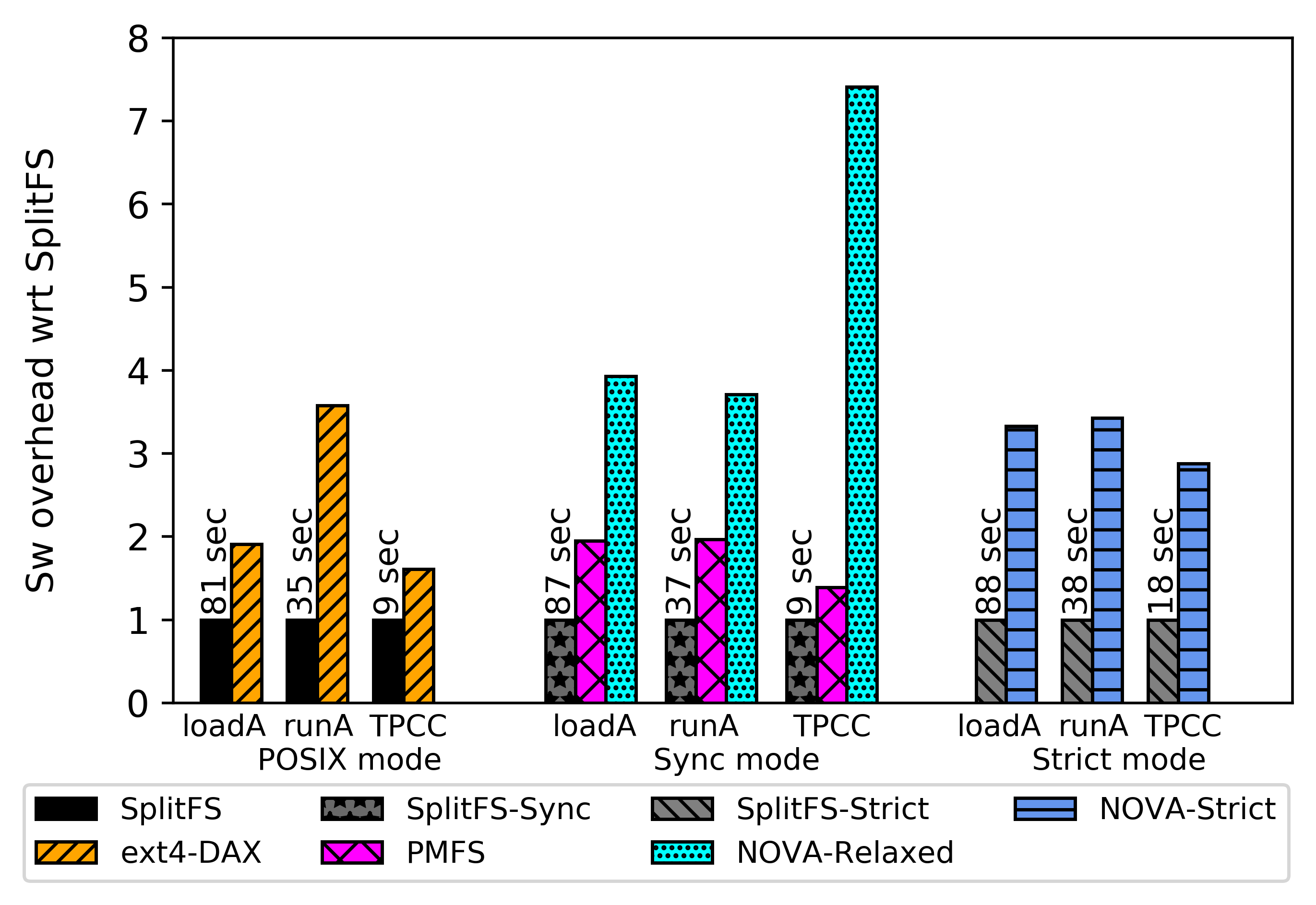}
\vspace{-20pt} \mycaption{Software overhead in applications}{ This
  figure shows the relative file system software overhead incurred by
  different applications with various file systems as compared
  \sysname providing the same level of consistency guarantees (lower
  is better). The numbers shown indicate the absolute time taken to
  run the workload for the baseline file system.}
\label{fig-sw-ovhd-apps}
\end{figure}

%% file: fig-real-apps.tex
\begin{figure*}
\centering
\includegraphics[width=\textwidth]{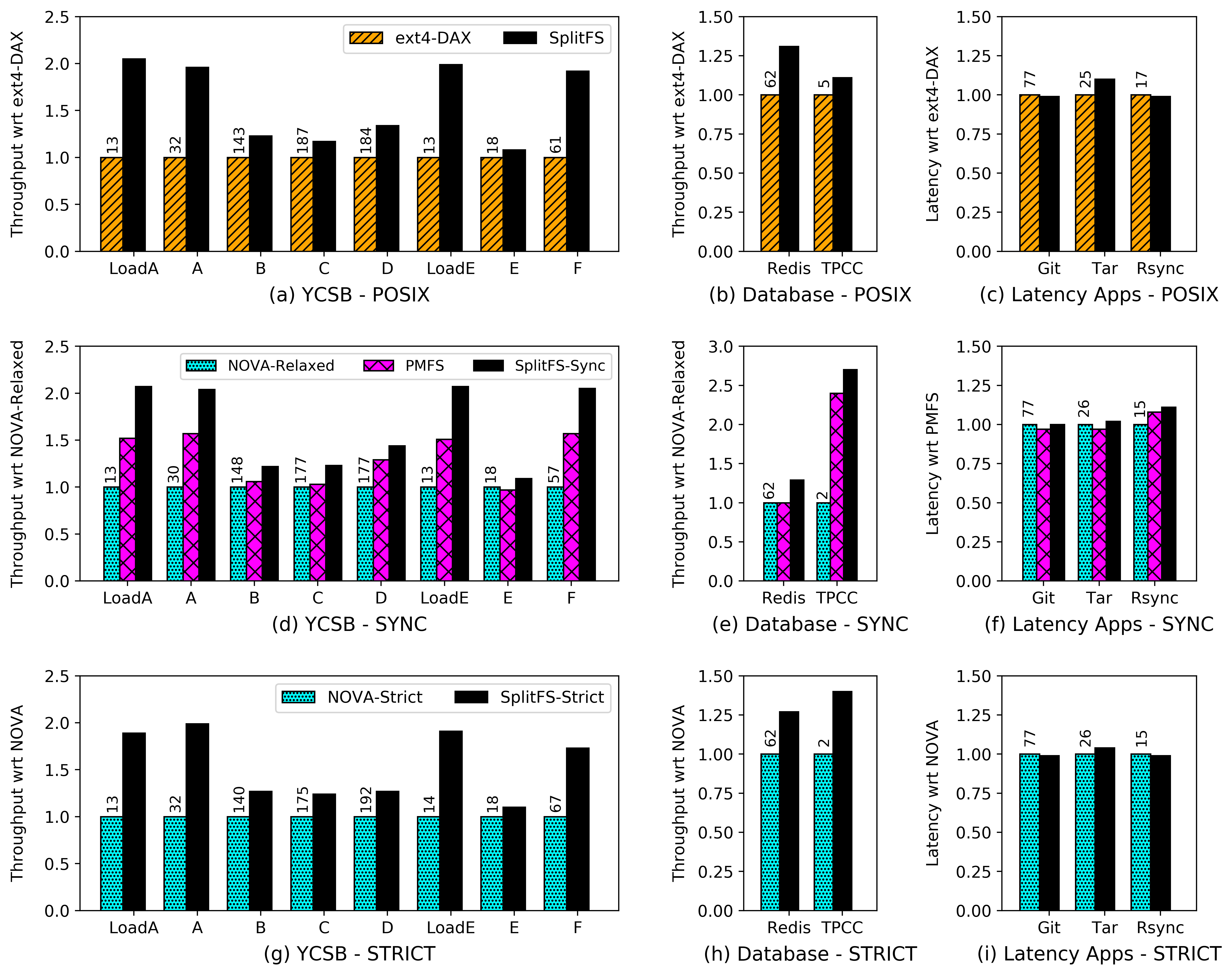}
\vspace{-20pt}

\mycaption{Real application performance}{This figure shows the
  performance of both data intensive applications (YCSB, Redis, and
  TPCC) and metadata instensive utilities (git, tar, and rsync) with
  different file systems, providing three different consistency
  guarantees, POSIX, sync, and strict. Overall, \sysname beats all
  other file systems on all data intensive applications (in their
  respective modes) while incurring minor performance degradation on
  metadata heavy workloads. For throughput workloads, higher is
  better. For latency workloads, lower is better. The numbers indicate
  the absolute throughput in Kops/s or latency in seconds for the base
  file system.}
\label{fig-real-apps}
\end{figure*}

%% file: tbl-splitfs-strata.tex
\begin{table}[!t]
  \small
  \centering
  \ra{1.3}
  \begin{tabular}{@{}p{80pt}p{80pt}p{30pt}@{}}
    \toprule[1.2pt]
    Workload & Strata & \sysname \\
    \midrule
    \addlinespace
    Load A & 29.1 kops/s & 1.73\myx \\
    Run A & 55.2 kops/s & 1.76\myx \\
    Run B & 76.8 kops/s & 2.16\myx \\
    Run C & 94.3 kops/s & 2.14\myx \\
    Run D & 113.1 kops/s & 2.25\myx \\
    Load E & 29.1 kops/s & 1.72\myx \\
    Run E & 8.1 kops/s & 2.03\myx \\
    Run F & 73.3 kops/s & 2.25\myx \\

    \bottomrule[1.2pt]
  \end{tabular}
  \vspace{3pt} \mycaption{\sysname vs. Strata}{This table compares
  the performance of Strata and \sysname strict running YCSB on LevelDB. We present the raw throughput numbers for 
  Strata and normalized \sysname strict throughput w.r.t Strata. This is the biggest workload that we could run reliably on Strata.}
  \label{tbl-splitfs-strata}
\end{table}

%% file: related.tex
\section{Related Work}
\label{sec-related}

\sysname builds on a large body of work on PM file systems and
building low-latency storage systems. We briefly describe the work
that is closest to \sysname.

\vheading{Aerie}. Aerie~\cite{aerie} was one of the first systems to
advocate for accessing PM from user-space. Aerie proposed a split
architecture similar to \sysname, with a user-space library file
system and a kernel component. Aerie used a user-space metadata server
to hand out leases, and only used the kernel component for
coarse-grained activities like allocation. In contrast, \sysname does
not use leases (instead making most operations immediately visible)
and uses ext4 DAX as its kernel component, passing all metadata
operations to the kernel. Aerie proposed eliminating the POSIX
interface, and aimed to provide applications flexibility in
interfaces. In contrast, \sysname aims to efficiently support the
POSIX interface.

\vheading{Strata}. The Strata~\cite{strata} cross-device file system
is similar to Aerie and \sysname in many respects. There are two main
differences from \sysname. First, Strata writes all data to a
process-private log, coalesces the data, and then writes it to a
shared space. In contrast, only appends are private (and only until
\vtt{fsync}) in \sysname; all metadata operations and overwrites are
immediately visible to all processes in \sysname. \sysname does not
need to copy data between a private space and a shared space; it
instead relinks data into the target file. Finally, since Strata is
implemented entirely in user-space, the authors had to re-implement a
lot of VFS functionality in their user-space library. \sysname instead
depends on the mature codebase of ext4 DAX for all metadata
operations.

\vheading{Quill and FLEX}. Quill~\cite{quill} and File Emulation with
DAX (FLEX)~\cite{flex} both share with \sysname the core technique of
transparently transforming read and overwrite POSIX calls into
processor loads and stores. However, while Quill and FLEX do not
provide strong semantics, \sysname can provide applications with
synchronous, atomic operations if required. \sysname also differs in
its handling of appends.  Quill calls into the kernel for every
operation, and FLEX optimizes appends by pre-allocating data beyond
what the application asks for. In contrast, \sysname elegantly handles
this problem using staging files and the relink primitive. While Quill
appends are slower than ext4 DAX, \sysname appends are faster than
ext4 DAX appends. At the time of writing this paper, FLEX has not been
made open-source, so we could not evaluate it.

\vheading{PM file systems}. Several file systems such as
SCMFS~\cite{scmfs}, BPFS~\cite{bpfs}, and NOVA~\cite{nova} have been
developed specifically for PM. While each file system tries to reduce
software overhead, they are unable to avoid the cost of trapping into
the kernel. The relink primitive from \sysname is similar to the
short-circuit paging presented in BPFS. However, while short-circuit
paging relies on an atomic 8-byte write, \sysname relies on ext4's
journaling mechanism to make relink atomic.

\vheading{Kernel By-Pass}. Several projects have advocated direct
user-space access to networking~\cite{unet}, storage~\cite{dafs,nas,
  petal, moneta}, and other hardware features~\cite{dune, ix,
  arrakis}. These projects typically follow the philosophy of
separating the control path and data path, as in
Exokernel~\cite{exokernel} and Nemesis~\cite{barham1997fs}. \sysname
follows this philosophy, but differs in the abstraction provided
by the kernel component; \sysname uses a PM file system as its
kernel component to handle all metadata operations, instead of limiting
it to lower-level decisions like allocation.

%% file: conc.tex
\section{Conclusion}
\label{sec-conc}

We present \sysname, a PM file system built using the split
architecture. \sysname handles data operations entirely in user-space,
and routes metadata operations through the ext4 DAX PM file
system. \sysname provides three modes with varying guarantees, and
allows applications running at the same time to use different
modes. \sysname only requires adding a single system call to the ext4
DAX file system. Evaluating \sysname with micro-benchmarks and real
applications, we show that it outperforms state-of-the-art PM file
systems like NOVA on many workloads. The design of \sysname allows
users to benefit from the maturity and constant development of the
ext4 DAX file system, while getting the performance and strong
guarantees of state-of-the-art PM file systems. \sysname is publicly available at \url{https://github.com/utsaslab/splitfs}.

%% file: ack.tex
\section*{Acknowledgments}
We would like to thank our shepherd, Keith Smith, the anonymous
reviewers, and members of the LASR group and the Systems and Storage
Lab for their feedback and guidance. We would like to thank Intel and
ETRI IITP/KEIT[2014-3-00035] for providing access to Optane DC
Persistent Memory for conducting experiments for the paper. This work
was supported by NSF CAREER grant 1751277 and generous donations from
VMware, Google and Facebook.
 Any opinions, findings, and conclusions,
 or recommendations expressed herein are those of the authors and do
 not necessarily reflect the views of other institutions.